\documentclass[aps,prl,fleqn,twocolumn,superscriptaddress,floatfix]{revtex4}

\usepackage{epsfig}
\usepackage{amsmath,amssymb,amsfonts}

\begin{document}

\title{Anomalous Viscosity of an Expanding Quark-Gluon Plasma}

\author{M.~Asakawa}
\affiliation{Department of Physics, Osaka University,
             Toyonaka 560-0043, Japan}

\author{S.~A.~Bass}
\affiliation{Department of Physics, Duke University, Durham, NC 27708}

\author{B.~M\"uller}
\affiliation{Department of Physics, Duke University, Durham, NC 27708}

\date{\today}

\begin{abstract}
We argue that an expanding quark-gluon plasma has an anomalous 
viscosity, which arises from interactions with dynamically
generated color fields. We derive an expression for the anomalous 
viscosity in the turbulent plasma domain and apply it to the 
hydrodynamic expansion phase, when the quark-gluon plasma is near
equilibrium. The anomalous viscosity dominates over
the collisional viscosity for weak coupling and not too late times. 
This effect may provide an explanation for the apparent ``nearly
perfect'' liquidity of the matter produced in nuclear collisions
at RHIC without the assumption that it is a strongly coupled state.
\end{abstract}

\maketitle

Measurements of the anisotropic collective flow of hadrons emitted
in noncentral collisions of heavy nuclei at the Relativistic Heavy 
Ion Collider (RHIC) are in remarkably good agreement with the 
predictions of ideal relativistic fluid dynamics \cite{RHIC_WP}.
In order to describe the data, calculations need to assume that the 
matter formed in the nuclear collision reaches thermal equilibrium
within a time $\tau_{\rm i} < 1$ fm/c \cite{Heinz:2001xi} and then 
expands with a very small shear viscosity $\eta \ll s$, where $s$ is 
the entropy density \cite{Teaney:2003pb}. The comparison between 
data and calculations indicates that the viscosity of the matter 
cannot be much larger than the postulated lower bound 
$\eta_{\rm min} = s/4\pi$ \cite{Kovtun:2004de}, which is reached in 
certain strongly coupled supersymmetric gauge theories 
\cite{Policastro:2001yc}.

This result is nontrivial because the shear viscosity of a weakly
coupled, perturbative quark-gluon plasma is not small. In fact, the
perturbative result for the shear viscosity, in leading logarithmic
approximation, is \cite{Arnold:2000dr}
\begin{equation}
\eta_{\rm C} = \frac{d_f T^3}{g^4 \ln g^{-1}} ,
\label{eq:eta-C}
\end{equation}
where $d_f \sim O(100)$ is a numerically determined constant that 
weakly depends on the number of quark flavors $n_f$. The result 
(\ref{eq:eta-C}), as well as the finding that numerical solutions of 
the Boltzmann equation exhibit fluid dynamical behavior only when 
the cross section between gluons is artificially increased by a 
large factor \cite{Molnar:2001ux}, have invited speculations 
that the matter produced at RHIC is a strongly coupled quark-gluon 
plasma (sQGP). The possible microscopic structure of such a state 
is not well understood at present 
\cite{Shuryak:2004tx,Koch:2005vg,Liao:2005pa}.

Here we present an alternative mechanism that may be responsible for 
a small viscosity of a weakly coupled, but expanding quark-gluon plasma.
The new mechanism is based on the theory of particle transport in 
turbulent plasmas \cite{Dupree:1966,Dupree:1968}. Such plasmas are 
characterized by strongly excited random field modes in certain regimes 
of instability, which coherently scatter the charged particles and 
thus reduce the rate of momentum transport. The scattering by turbulent 
fields in electromagnetic plasmas is known to greatly increase the 
energy loss of charged particles \cite{Okada:1980} and reduce the 
heat conductivity \cite{Malone:1975,Okada:1978} and the viscosity 
\cite{Abe:1980a,Abe:1980b} of the plasma. Following Abe and Niu 
\cite{Abe:1980b}, we call the contribution from turbulent fields 
to transport coefficients ``anomalous.''

The sufficient condition for the spontaneous formation of turbulent, 
partially coherent fields is the presence of instabilities in 
the gauge field due to the presence of charged particles. This 
condition is met in electromagnetic plasmas with an anisotropic 
momentum distribution of the charged particles \cite{Weibel:1959}, 
and it is known to be satisfied in quark-gluon plasmas with an 
anisotropic momentum distribution of thermal partons 
\cite{Mrowczynski:1988dz,Mrowczynski:1993qm,Romatschke:2003ms}.

Most of the work exploring the consequences of the instabilities 
\cite{Arnold:2003rq,Randrup:2003cw,Mueller:2005un} has been focused 
on the early stage of the collision, when the momentum distribution 
is highly anisotropic and far from equilibrium. It was pointed out 
that the fields generated by the instabilities will drive the parton
distribution rapidly toward local isotropy and thus into the 
hydrodynamical regime \cite{Arnold:2004ti}. Here we are concerned 
with the later stage of the reaction, when the matter is nearly 
equilibrated and evolves by hydrodynamical expansion. Because the 
partonic plasma expands rapidly, the momentum distribution of the
partons remains anisotropic even at late times, with the size of the
anisotropy being proportional to the viscosity. 

As we will show, the turbulent plasma fields induce an additional,
anomalous contribution to the viscosity, which we denote as $\eta_{\rm A}$. 
This anomalous viscosity decreases with increasing strength of the 
turbulent fields. Since the amplitude of the turbulent fields grows 
with the magnitude of the momentum anisotropy, a large anisotropy 
will lead to a small value of $\eta_{\rm A}$. Because the relaxation rates 
due to different processes are additive, the total viscosity is given by 
\begin{equation}
\eta^{-1} = \eta_{\rm A}^{-1} + \eta_{\rm C}^{-1} .
\label{eq:eta-total}
\end{equation}
This equation implies that $\eta_{\rm A}$ dominates the total shear viscosity, 
if it is smaller than $\eta_{\rm C}$. In that limit, the anomalous mechanism
exhibits a stable equilibrium in which the viscosity regulates itself: 
The anisotropy grows with $\eta$, but an increased anisotropy tends to 
suppress $\eta_{\rm A}$ and thus $\eta \approx \eta_{\rm A}$. We derive the 
resulting self-consistency condition for $\eta_{\rm A}$ below.

The fireballs formed in relativistic heavy ion 
collisions exhibit collective expansion in both, the longitudinal 
and the transverse directions with respect to the beam axis. Here 
we focus on the longitudinal expansion, but our arguments apply 
as well to the transverse expansion component. We assume
the longitudinal flow profile during the hydrodynamic expansion
phase of a relativistic heavy ion collision to be approximately boost 
invariant \cite{Bjorken:1982qr} and of the form $u_z(z,t)=z/t$, 
where $z$ and $t$ are measured from the collision point in the 
center-of-mass frame. The velocity gradient $\partial u_z/\partial z$ 
leads to an anisotropy in the local momentum distribution  
\cite{Heiselberg:1995sh,Teaney:2003pb} 
\begin{equation}
\frac{2T_{zz}}{T_{xx}+T_{yy}} - 1
= - \frac{8}{T\tau} \frac{\eta}{s} ,
\label{eq:aniso}
\end{equation}
where $T$ denotes the temperature, $s$ the entropy density of the 
matter, and $\tau=\sqrt{t^2-z^2}$ is the time in local comoving 
coordinates. For simplicity, we have assumed that the equation of 
state of the matter is that of free massless partons,
$\epsilon=3P=3sT/4$. As (\ref{eq:aniso}) shows, the anisotropy is 
linearly dependent on the viscosity of the matter.

As already mentioned, any anisotropy of the local momentum distribution
of quasi-thermal partons engenders instabilities of soft gluon modes
in the momentum regime $k<gT$, whose growth rate increases in proportion 
to the anisotropy \cite{Romatschke:2003ms}. In electromagnetic plasmas 
the growth of the instability eventually saturates when the increasing 
field modifies the particle distribution in such a way that the instability 
is eliminated. In nonabelian plasmas the nonlinear self-interactions of 
the gauge field restrict the growth of the unstable modes; this mechanism
dominates at weak coupling \cite{Rebhan:2004ur,Rebhan:2005re,Arnold:2005vb}. 
In the quasi-stationary state that is reached in the nonlinear domain, 
energy absorbed from the thermal partons cascades from the most unstable 
gauge field modes into modes of increasingly shorter wavelengths. The 
power spectrum of this energy cascade has the form $P(k)\sim k^{-2}$, 
analogous to the Kolmogorov cascade in a turbulent fluid \cite{Arnold:2005ef}. 

As already stated, we are here concerned with the effect of the 
unstable field modes on the transport properties 
of the medium when it has reached the collective expansion phase.
Following the standard Chapman-Enskog theory of transport coefficients, 
we assume that the plasma is driven only slightly out of 
equilibrium by the collective flow and that the local phase-space 
distribution can be written as
\begin{equation}
f(\vec{p},\vec{r}) 
= f_0(\vec{p}) [1+f_1(\vec{p},\vec{r})(1\pm f_0(\vec{p}))] ,
\label{eq:f0-f1}
\end{equation}
where $f_0(\vec{p})$ is the local equilibrium distribution, $+(-)$ 
applies to bosons (fermions), and $f_1$ has the form 
\begin{equation}
f_1(\vec{p},\vec{x}) 
= - \frac{\bar\Delta}{2E T^2} p_ip_j \Delta_{ij}(u) .
\label{eq:f1}
\end{equation}
Here $\Delta_{ij}(u) = (\nabla_iu_j+\nabla_ju_i 
  -\frac{2}{3}\delta_{ij}\nabla\cdot\vec{u})$ 
is the traceless part of the flow gradient related to the shear 
viscosity, and $\bar\Delta$ parametrizes the strength of the anisotropy.
For massless particles, $\bar\Delta$ is related to the macroscopic transport 
coefficient of shear viscosity by $\bar\Delta = 5\eta/s$.
The boost invariant longitudinal expansion corresponds to the 
perturbation
\begin{equation}
f_1(\vec{p}) = - \frac{\bar\Delta}{3E T^2\tau} (3p_z^2-p^2) .
\label{eq:f1-Qzz}
\end{equation}

In order to explore the response of the plasma to this perturbation, 
we need to determine the influence of the saturated field modes on 
the propagation of thermal plasma particles. The relevant transport 
theory was developed by Dupree \cite{Dupree:1966,Dupree:1968} for 
an electromagnetic plasma in the limit of strong turbulence and weak 
coupling. We now generalize this formalism to a nonabelian plasma. 
Our starting point is the Vlasov-Boltzmann equation for the phase 
space distribution of color charges $Q^a$ in a color-magnetic field
$\vec{B}^a$:
\begin{equation}
\left[ \frac{\partial}{\partial t} + \vec{v}\cdot\nabla_r 
+ \vec{F}\cdot\nabla_p \right] f(\vec{r},\vec{p},t) = C[f] ,
\label{eq:Vlasov}
\end{equation}
where $\vec{v}=\vec{p}/E$ is the velocity of a thermal parton with 
momentum $\vec{p}$ and energy $E$, 
$\vec{F}=g Q^a (\vec{E}^a + \vec{v}\times\vec{B}^a)$ is the color
Lorentz force, and $C[f]$ denotes the collision term. We focus here 
on the effects of the Vlasov term and refer to the results of Arnold, 
Moore, and Yaffe \cite{Arnold:2000dr} for the viscosity due to incoherent 
collisions. Because transverse field modes have the highest growth rates 
in the linear regime and transverse color-electric fields are less 
effective in restoring the particle distribution to isotropy, we here 
concentrate on the effects due to coherent color-magnetic field modes. 
The effects of longitudinal electric field modes, which are strongly 
excited in the nonlinear domain and can also lead to isotropy, will be 
discussed in a forthcoming longer publication \cite{ABM_tbp}.

In order to isolate the dissipative effects of the color field, one 
averages the particle trajectories over an ensemble of color-magnetic 
fields. Assuming that $\langle B^a\rangle=0$ and factorizing higher 
than second moments of the field distribution, one can then show that 
the ensemble averaged phase space distribution $\bar{f}$ satisfies an 
equation of the Fokker-Planck type \cite{Dupree:1966}:
\begin{equation}
\left[ \frac{\partial}{\partial t} + \vec{v}\cdot\nabla_r 
- \nabla_p D(\vec{p},t)\nabla_p \right] \bar{f} = C[\bar{f}]  
\label{eq:F-P}
\end{equation}
with the diffusion tensor
\begin{equation}
D_{ij} = \int_{-\infty}^t dt' \left\langle 
  F_i(\bar{r}(t'),t') F_j(\bar{r}(t),t) \right\rangle .
\label{eq:D-def}
\end{equation}

Dupree's treatment \cite{Dupree:1966} is based on the argument that the 
Fourier components of the coherent field, $B(k,\omega_k)$ are slowly 
varying functions and the autocorrelation function of the Lorentz 
force in (\ref{eq:D-def}) is determined by the action of the magnetic 
field on the particle trajectories. This leads to a self-consistency 
condition for the mean deviation $\langle\Delta\bar{r}^2\rangle$ of 
the particle trajectories from straight lines. Because the growth 
and coherence of the nonabelian gauge field (at weak coupling) is 
not controlled by the back reaction of the distorted particle 
trajectories on the gauge field, but by the inherent nonlinearities
of the gauge field itself \cite{Arnold:2003rq}, we here adopt a 
different approach. The velocity of 
propagation of the collective modes of the nonabelian plasma is 
less than the speed of light, while the thermal particles move 
(nearly) at the speed of light. The autocorrelation function of 
the color-magnetic field along the path of a particle will thus 
be controlled by the spatial correlation length of the fields created 
by the growth of the unstable modes. Assuming that the correlation 
length for the color-magnetic fields is short in comparison with 
the curvature radius of the trajectory of a plasma particle, we 
can then take $\bar{v}(t)=\bar{v}(t')=v$ out of the average in 
(\ref{eq:D-def}) and are left with the autocorrelation function 
of the magnetic field along a typical particle trajectory:
\begin{equation}
\int_{-\infty}^t dt'\, \langle B^a_i(t') B^b_j(t) \rangle 
\equiv \langle B^a_i B^b_j \rangle\, \tau_{\rm m} .
\label{eq:tau-mem}
\end{equation}
In our case, the color-magnetic fields generated by the plasma 
instability point in a transverse direction with respect to the beam. 
Assuming that the ensemble average is diagonal in color and employing 
the notation $\vec{L}^{(p)}= -i\vec{p}\times\nabla_p$, we can write 
the diffusive term as
\begin{equation}
\nabla_p D(p) \nabla_p 
= - \frac{g^2 Q^2}{2(N_c^2-1)E^2} \langle B^2 \rangle\, \tau_{\rm m} 
    (L^{(p)}_{\perp})^2 ,
\label{eq:D-1}
\end{equation}
where $N_c=3$ is the number of colors and the index $\perp$ denotes
the components transverse to the beam axis.

The action of the diffusion operator on $\bar{f}$ is easily evaluated by
noting that the perturbation $f_1(\vec{p})$ has quadrupole form.
In order to derive the anomalous viscosity due to the diffusion term,
we follow Abe and Niu \cite{Abe:1980b} and take moments of the drift
and diffusion terms in (\ref{eq:F-P}) with $p_z^2$. Using massless 
quarks and gluons in the momentum integrals, we obtain:
\begin{equation}
\int \frac{d^3p}{(2\pi)^3E}\, p_z^2 \vec{v}\cdot\nabla_r \bar{f}(\vec{p})
= \frac{1}{T\tau} \frac{16\nu_4\zeta(4)}{15\pi^2} T^5  ,
\label{eq:mom-drift}
\end{equation}
\begin{eqnarray}
\int \frac{d^3p}{(2\pi)^3E} p_z^2 
     \nabla_p D(p) \nabla_p \bar{f}(\vec{p}) \hspace{1truein}
\nonumber \\
\qquad = \frac{1}{T\tau} 
    \frac{\bar\Delta\, g^2 \langle B^2 \rangle \tau_{\rm m} }{(N_c^2-1)} 
    \frac{4N_c\nu'_2\zeta(2)}{15\pi^2} T^2 ,
\label{eq:mom-diffus}
\end{eqnarray}
where 
\begin{eqnarray}
\nu_N &=& 16+12(1-2^{-N})n_f ,
\\
\nu'_N &=& 16+6(1-2^{-N})n_f(N_c^2-1)/N_c^2 ,
\end{eqnarray}
 and we have used that $Q^2=N_c$ for gluons and $Q^2=(N_c^2-1)/(2N_c)$ 
for quarks. Equating the two results and using the relation 
$\bar\Delta=5\eta/s$ we obtain the sought after expression for the 
anomalous shear viscosity due to the coherent color-magnetic fields:
\begin{equation}
\eta_{\rm A} = \frac{4(N_c^2-1)\nu_4\zeta(4)}{5N_c\nu'_2\zeta(2)} 
         \frac{s T^3}{g^2\langle B^2 \rangle\,\tau_{\rm m}} .
\label{eq:eta-B}
\end{equation}
It is noteworthy that the rhs. of (\ref{eq:eta-B}) itself depends 
implicitly on the viscosity, because the intensity of the turbulent 
fields grows with increasing anisotropy of the momentum distribution 
in the plasma. 

Next we need to address the question how large $\langle B^2 \rangle$
and $\tau_{\rm m}$ are. The coherent color magnetic fields are only
generated by the plasma instability when the momentum distribution 
of partons in the quark-gluon plasma is deformed due to the collective
expansion. We know from analytical studies how the growth rate of the 
instability depends on the anisotropy of the momentum distribution, 
but there are no published systematic studies that show how the 
saturation level of the coherent field energy depends on the anisotropy. 

The study by Romatschke and Strickland \cite{Romatschke:2003ms} expresses 
the anisotropy in terms of a parameter $\xi$ and a unit vector $\hat{n}$.
Choosing $\hat{n}=\hat{e}_z$, this {\em ansatz} corresponds to a 
perturbation of the equilibrium distribution of the form (\ref{eq:f1-Qzz})
with $\bar\Delta=\xi T\tau/2$.
The average intensity of the coherent color-magnetic fields is a 
function of the momentum anisotropy. For lack of a precise knowledge
of this function, we here parametrize it as a power law:
$g^2 \langle B^2 \rangle = b_0 g^4 T^4 \xi^{b_1}$,
and conjecture a linear dependence ($b_1=1$). We argued above that 
the scale for the memory time $\tau_{\rm m}$ for the nonabelian plasma 
will be set by the spatial coherence length of the coherent fields. 
This coherence length is given by the wavelength of the maximally 
unstable mode, which is of the order of the Debye length:
$\tau_{\rm m} \sim \mu_D^{-1} \sim (gT)^{-1}$,
where the last form assumes weak coupling. We currently lack a precise 
determination of $\tau_{\rm m}$, which could be obtained from  
numerical solutions of the Yang-Mills equations for the anisotropy 
discussed here.

Using the relations between $\xi$, $\bar\Delta$, and $\eta$, we can 
now state the dependence of the rhs.\ of (\ref{eq:eta-B}) on the viscosity:
\begin{equation}
g^2\langle B^2 \rangle\, \tau_{\rm m} 
= 10 b_0 g^3 \frac{T^2\eta}{s \tau} .
\label{eq:g2B2t}
\end{equation}
Note that $\eta$ here is the viscosity due to all sources of 
dissipation, including collisions. If we neglect the viscosity 
$\eta_{\rm C}$ due to particle collisions and set $\eta=\eta_{\rm A}$ on the 
right-hand side of (\ref{eq:g2B2t}), eq.~(\ref{eq:eta-B}) yields 
a self-consistency condition for the anomalous viscosity $\eta_{\rm A}$,
which has the solution:
\begin{equation}
\eta_{\rm A} = 
\left(\frac{2(N_c^2-1)\nu_4\zeta(4)T\tau}
           {25b_0N_c\nu'_2\zeta(2)}\right)^{1/2}
       \frac{s}{g^{3/2}} .
\label{eq:etaB-1}
\end{equation}

Several things are notable about this result. First, if the memory
time $\tau_{\rm m}$ is longer than our estimate, the 
value of $\eta_{\rm A}$ decreases. Second, the dependence of $\eta_{\rm A}$
on the gauge coupling is parametrically much weaker than that
of the collisional viscosity (\ref{eq:eta-C}). Thus, for weak coupling 
$g\ll 1$ and not too late times $\tau$, the anomalous viscosity 
will be much smaller than the collisional viscosity. According to
(\ref{eq:eta-total}), this implies that $\eta_{\rm A}$ is dominant at early 
times, and the collisional viscosity $\eta_{\rm C}$ may dominate at large 
times. The cross-over time $\tau_c$ between the two regimes is given 
by the condition $\eta_{\rm A}=\eta_{\rm C}$. 
Because of the failure of the perturbative result for $\eta_{\rm C}$ in
the domain $g \geq 1$, it is difficult to give a reliable estimate
of the cross-over time between the two regimes. Ignoring those 
limitations and assuming $b_0 \sim O(1)$, one can surmise that the
anomalous viscosity dominate for a few fm/c in heavy-ion collisions
at RHIC or LHC energies. 
At late times, the transverse expansion of the medium also needs to
be taken into account. Assuming a radial dependence of the form
$u_r(r)\approx\beta_0 r/R$, where $R$ is the nuclear radius, the 
anisotropy due to transverse flow is
\begin{equation}
p_ip_j \Delta_{ij}(u) = - \frac{2\beta_0}{3R}(3p_z^2-p^2) .
\label{eq:transv}
\end{equation}
Comparing with (\ref{eq:f1-Qzz}) one sees that the effects from the 
radial expansion become comparable to those from the longitudinal 
expansion when $\beta_0 \sim R/\tau$. The anomalous contribution 
to the viscosity may, therefore, never be negligible during the 
life-time of the plasma phase. We note that collisions among thermal 
particles may suppress the Weibel instability when the coupling constant 
$\alpha_s$ exceeds a threshold value \cite{Schenke:2006xu}. It is 
presently unknown whether this occurs before or after the cross-over 
between $\eta_{\rm A}$ and $\eta_{\rm C}$ for experimentally relevant 
conditions.

The approach outlined here can be used to derive other anomalous 
transport properties of an expanding, turbulent quark-gluon plasma.
Maybe the most important among these are the coefficient $\hat{q}$ 
of radiative energy loss of an energetic parton \cite{Baier:2000mf},
which might be increased by scattering on turbulent fields. 

In summary, we have shown that an expanding quark-gluon plasma 
acquires an anomalous viscosity due to the interaction of thermal
partons with chromomagnetic fields generated by instabilities of soft 
field modes. In the weak coupling limit, the anomalous viscosity is 
much smaller than the viscosity due to collisions among thermal partons. 
By reducing the shear viscosity of a weakly coupled, but expanding 
quark-gluon plasma, this mechanism could possibly explain the 
observations of the RHIC experiments without the assumption of a 
strongly coupled plasma state. A definitive answer will require the 
numerical evaluation of the correlation function (\ref{eq:tau-mem})
as a function of the anisotropy parameter $\bar\Delta$.


{\it Acknowledgments}: This work was supported in part by  
grants from the U.~S.~Department of Energy (DE-FG02-05ER41367), 
the National Science Foundation (NSF-INT-03-35392), and the 
Japanese Ministry of Education (grant-in-aid 17540255).

\end{document}